\documentclass[1p]{elsarticle-self}

\usepackage{hyperref}

\usepackage{amsmath}
\usepackage{amsfonts}
\usepackage{bm}
\usepackage{booktabs}
\usepackage[subrefformat=parens]{subcaption}
\usepackage{color}

\biboptions{authoryear}

\begin{document}

\begin{frontmatter}

\title{Mitigating Cognitive Biases in \\ Multi-Criteria Crowd Assessment}

\author[ku]{Shun Ito}

\author[ku]{Hisashi Kashima}

\address[ku]{Kyoto University, Yoshida-Honmachi, Sakyo-ku, Kyoto, 606-8501, Japan}

\begin{abstract}
Crowdsourcing is an easy, cheap, and fast way to perform large scale quality assessment; however, human judgments are often influenced by cognitive biases, which lowers their credibility. 
In this study, we focus on cognitive biases associated with a multi-criteria assessment in crowdsourcing; crowdworkers who rate targets with multiple different criteria simultaneously may provide biased responses due to prominence of some criteria or global impressions of the evaluation targets. 
To identify and mitigate such biases, we first create evaluation datasets using crowdsourcing and investigate the effect of inter-criteria cognitive biases on crowdworker responses. 
Then, we propose two specific model structures for Bayesian opinion aggregation models that consider inter-criteria relations. 
Our experiments show that incorporating our proposed structures into the aggregation model is effective to reduce the cognitive biases and help obtain more accurate aggregation results. 
\end{abstract}


\end{frontmatter}

\section{Introduction}\label{introduction_section}
Quality \textcolor{black}{assessments} of \textcolor{black}{various} objects, such as texts, images, and even customer services, plays \textcolor{black}{a direct and indirect} important role in many applications, and is often based on human subjectivity. 
Despite recent advances in AI and machine learning technologies, many applications still require human assessment because the characteristics of objects that can explain human subjectivity are sometimes unknown or too vague to be extracted automatically, which is a serious bottleneck \textcolor{black}{when conducting} large-scale automated quality \textcolor{black}{assessments}. 
The use of crowdsourcing is a promising way to implement this with the wisdom of the crowd.

\begin{figure}[tb]
 \begin{center}
  \includegraphics[width=8cm]{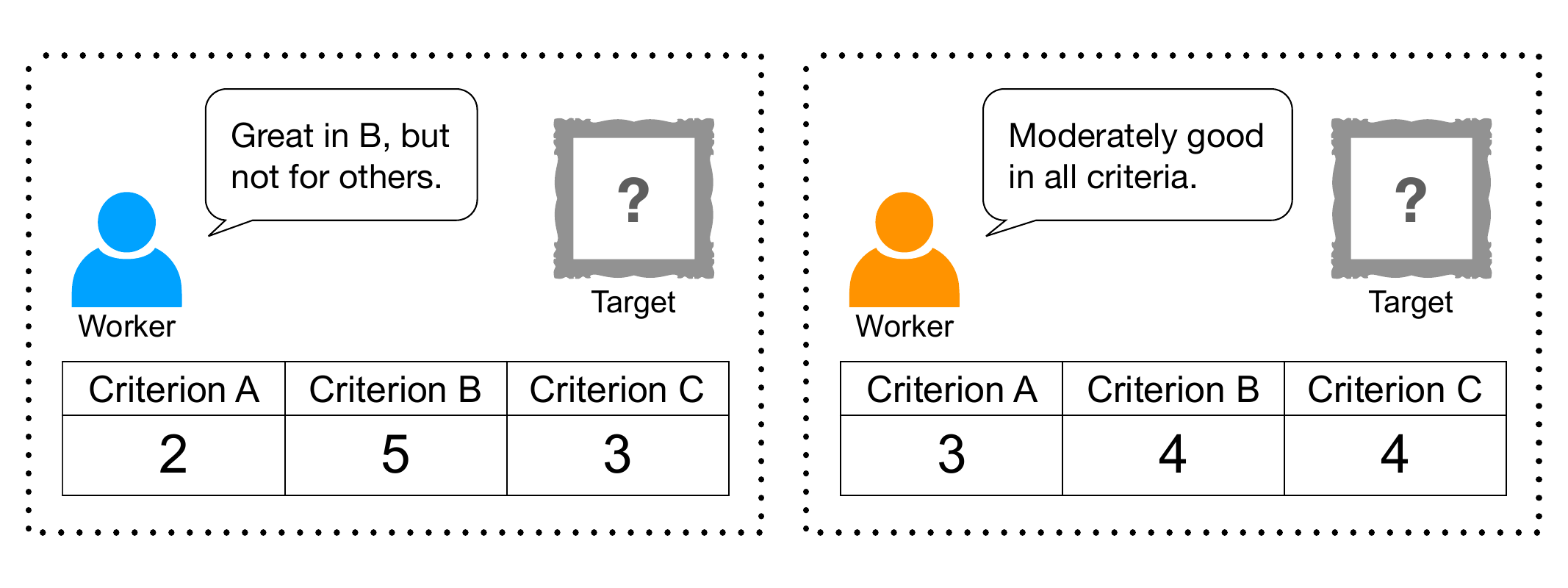}
  \subcaption{Unbiased examples. }
  \label{figure1a}
 \end{center}
 \begin{center}
  \includegraphics[width=8cm]{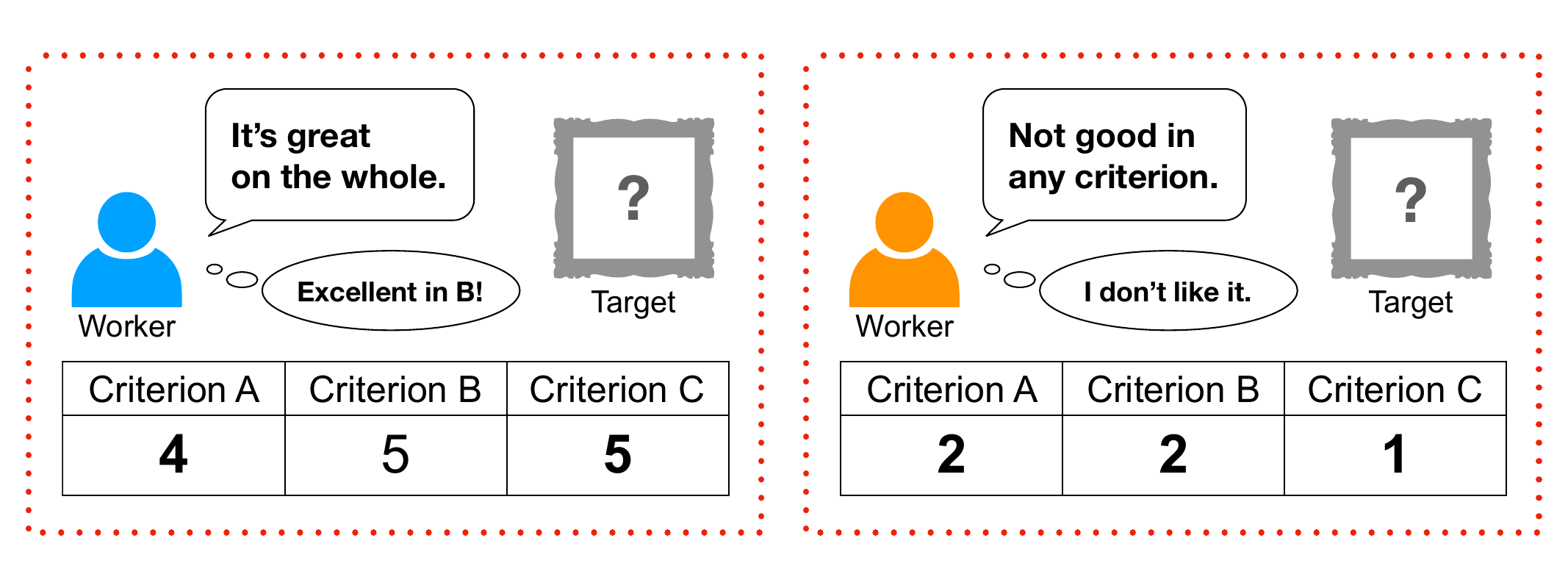}
  \subcaption{Biased examples. }
  \label{figure1b}
 \end{center}
 \caption{Two example cases of inter-criteria cognitive biases in multi-criteria \textcolor{black}{assessments}. The left is the case in which the worker rates Criterion B high, and therefore over-estimates the other two criteria affected by Criterion B. \textcolor{black}{On} the right the worker rates all criteria lower than unbiased evaluations due to the negative global impression of the target.}
 \label{figure1}
\end{figure}

One challenge in crowdsourced quality \textcolor{black}{assessments} is the uncertainty of human judgments. 
\textcolor{black}{Since} workers have different competences, expertise, or motivations, their responses are sometimes too noisy to analyze and extract useful knowledge.
A straightforward solution is to assign multiple crowdworkers to each evaluation target and aggregate the redundantly collected evaluations using majority voting.
More sophisticated statistical methods, such as Bayesian generative models, have also been explored for better aggregations. 
Various factors of human error have been introduced into statistical models, such as \textcolor{black}{the} ability of workers~\citep{dawid1979}, difficulty of \textcolor{black}{the} questions~\citep{whitehill2009,welinder2010}, and presence of malicious workers~\citep{raykar2011}. 

Psychologists have investigated the perceptual effects in decision-making processes of humans called {\it cognitive biases}.
Cognitive biases are systematic patterns of deviation from norms or rationality in human judgments \textcolor{black}{and} can potentially influence all types of human responses, including quality assessment.
Recent crowdsourcing studies \textcolor{black}{investigated} the existence of various cognitive biases and their effects on crowdworkers' responses.
For example, Eickhoff introduced four common cognitive biases encountered in crowd annotation and measured the magnitude of their impact on responses and machine learning-based predictions~\citep{eickhoff2018}.
Although other researchers have also investigated several cognitive biases~\citep{newell2016,gadiraju2017}, there is still a gap between the vast variety of cognitive biases studied in psychology and crowdsourcing research.

In the present study, we raise the issue of inter-criteria cognitive biases in crowdsourced quality \textcolor{black}{assessments}.
Inter-criteria cognitive biases such as the {\it halo error} are the evaluator biases that can occur when dealing with multiple perspectives simultaneously, \textcolor{black}{and} has been studied for years in the field of psychology~\citep{thorndike1920,decoths1977,balzer1992}.
In crowdsourcing, the multi-criteria assessment that takes into account multiple attributes of the target object and scores each criterion\textcolor{black}{,} is a general form of quality assessment~\citep{manouselis2007,Lester1997}.
\textcolor{black}{More} monetary and time costs \textcolor{black}{are incurred when assessing multiple criteria than when} assessing a single criterion; thus, a more practical choice would be to conduct \textcolor{black}{the} assessments for all criteria at once as a single crowdsourcing task.
Although such task design could lead to inter-criteria cognitive biases, their impact and mitigation in crowdsourcing has not yet been fully discussed.

In inter-criteria cognitive bias situations, crowdworkers give lenient responses or similar ratings across criteria.
Figure~\ref{figure1} shows schematic examples of biased judgments in multi-criteria \textcolor{black}{assessments}.
One example is \textcolor{black}{where} workers \textcolor{black}{notice} significantly high or low quality in one of the evaluation criteria; they may over- or under-estimate other criteria by being influenced by the significant criteria.
Another example is the case in which workers judge each criterion based on the overall impression of the target itself; the variance of their responses across the different criteria could be smaller due to the strong emphasis on the overall impression. 
Such biased responses can also negatively impact the predictive performance of the evaluation aggregation models.

We first investigate the presence of such cognitive biases in multi-criteria assessment from real data, then consider statistical generative models for aggregating multi-criteria assessments with novel model structures that consider the effect of inter-criteria cognitive biases.
\color{black}
First, we construct two datasets of short text writings as our evaluation targets, and then collect responses with multi-criteria quality ratings for each target using crowdsourcing. 
To investigate the inter-criteria effect, we collect the evaluations in two different task settings: (i) evaluating different criteria in separate tasks and (ii) simultaneously evaluating all criteria in a single task. 
We treat the responses in the first setting as unbiased responses, and those in the second setting as biased responses. 
We perform statistical analysis to compare these two settings and reveal how inter-criteria biases bias crowdworkers' judgments in terms of mean and variance. 
Second, based on the observation, we introduce a simple Bayesian aggregation model that predicts the true quality of the evaluation targets from the collected ratings, and then propose two specific model structures that extend the baseline model to mitigate the effects of inter-criteria biases. 
Finally, we conduct experiments using our datasets with varying numbers of workers. 
Our experimental results demonstrate that our proposed models outperform the baseline model in making debiased predictions.

In summary, the contributions of this study are three-fold. 
\begin{itemize}
    \item We identify inter-criteria cognitive biases in multi-criteria assessment in crowdsourcing.
    \item We construct real multi-criteria assessment datasets using crowdsourcing and demonstrate the effect of inter-criteria cognitive biases on the worker responses.
    \item We propose model structures to mitigate the inter-criteria effect and show their advantages in multi-criteria assessment aggregation.
\end{itemize}

The rest of the paper is organized as follows.
Section~\ref{dataset_section} describes the procedure of creating a real multi-criteria evaluation dataset using crowdsourcing; it then reveals the existence of inter-criteria cognitive biases caused by evaluating multiple criteria at once.
Section~\ref{proposedmethod_section} introduces a standard multi-criteria rating aggregation model and we extend the model to consider the inter-criteria effect.
Section~\ref{experiments_section} explains the experimental settings and shows the results of our proposed models.
Section~\ref{conclusion_section} concludes the paper and presents future directions.

\section{Related Work}\label{relatedwork_section}
One of the primary concerns with crowdsourcing human assessments is quality control.
Redundant execution of assessment tasks by multiple annotators is one of the most generic ways to improve the reliability of crowdsourced responses, and various statistical response aggregation methods have been explored.
\textcolor{black}{Originating} from the seminal study by Dawid and Skene~\citep{dawid1979} that aggregates responses from different annotators, researchers have extended statistical aggregation models by including various factors of annotation errors, for example, task difficulty~\citep{whitehill2009,welinder2010}, worker similarity~\citep{Venanzi2014}, and item similarity~\citep{lakkaraju2015}.
While they focused on categorical annotation, several studies have considered ordinal or numerical ratings, and aggregated worker responses based on the graded response model~\citep{raykar2010,baba2013} or the normal distribution~\citep{nguyen2016}.
In recent years, various label aggregation methods based on Bayesian generative models and Bayesian inference have been proposed to \textcolor{black}{address} a small number of worker labels.
For example, Bayesian Classifier Combination (BCC) is a Bayesian extension of the Dawid-and-Skene model, and Community BCC~\citep{Venanzi2014} and Enhanced BCC~\citep{Li2019-EBCC} are further extensions of \textcolor{black}{the} BCC \textcolor{black}{that} consider group structures of similar workers and correlated behaviors of workers, respectively.

Ability is not the only human factor that affects response quality, and various psychological and cognitive factors exist.
A large amount of psychological research has investigated cognitive biases related to rating tasks such as the leniency and halo errors~\citep{saal1980}.
The halo error \textcolor{black}{is} an inter-criteria effect where raters' inability or unwillingness to distinguish multiple dimensions of a given task~\citep{thorndike1920,decoths1977,balzer1992} \textcolor{black}{has been investigated}.

Inspired by the above cognitive psychological considerations, researchers in the crowdsourcing community have also investigated various types of biases that \textcolor{black}{reduce} the reliability of datasets.
Newell and Ruths showed that responses of repeatedly participating workers on image labeling tasks are biased by their previous tasks~\citep{newell2016}.
Eickhoff has systematically investigated four types of cognitive effects---ambiguity, anchoring, bandwagon, and decoy effect---based on experimental studies on crowdsourced document relevance assessment tasks~\citep{eickhoff2018}.
The above mentioned halo error is also observed in crowdsourced annotations such as human impression~\citep{biel2014} and peer assessment~\citep{kulkarni2014}.
Recent studies discussed confirmation biases that are derived from annotators' individual view or their background~\citep{barbera2020,coscia2020}.
Another line of research has considered biases related to task-design, such as the study of Kamar {\it et al.} who introduced the task-dependent bias in aggregation models~\citep{kamar2015}.

Attempts to reduce various cognitive psychological biases have also been made in the context of crowdsourcing research.
Typical approaches for bias mitigation are strategic task design or interventions in the annotation process.
Eickhoff and Vries explored several factors in crowdsourcing tasks that reduce cheat submissions~\citep{eickhoff2013}.
Faltings {\it et al.} showed that monetary incentives affect worker biases and proposed a game-theoretic bonus scheme~\citep{newell2016}.
Gadiraju {\it et al.} prepared pre-screening tasks where workers submit their self-assessment as well as their responses, which \textcolor{black}{aimed} to find competent workers based on the Dunning-Kruger effect~\citep{gadiraju2017self}.
Hube {\it et al.} focused on subjective judgments and found that instructing workers to consider social opinions or avoid involving their own potential biases\textcolor{black}{,} can actually mitigate biases stemming from strong opinions~\citep{hube2019}.

The existing research that is closest to our current study is about reducing certain cognitive biases in the post hoc integration of crowdsourced responses.
\cite{zhuang2015kdd} and \cite{zhuang2015wsdm} proposed Bayesian models for mitigating in batch annotation biases.
They assume that workers who work on multiple items in a single task balance the proportion of positive labels with that of negative labels among items~\citep{zhuang2015wsdm}, or respond based on the implicit ranking of items~\citep{zhuang2015kdd}.
Our study focuses on multi-criteria assessment where workers evaluate an item in multiple criteria in a single task.
Inspired by psychological research, we introduce the leniency and halo errors to crowdsourcing, where we assume that significant criteria or workers' impressions of evaluation \textcolor{black}{objects} could lead to inter-criteria biases.
We aim to mitigate such biases by extending an aggregation model and improve prediction performance of the quality of evaluation \textcolor{black}{objects}.
\color{black}

\begin{table*}[t]
 \centering
 \caption{
   {Comparison of the statistical generative models for aggregating crowdsourced annotations by the types of biases they consider. 
   Our proposed models ({\sc ImpCIM/CDM/ImpCDM}) include the cognitive bias as well as the worker and item biases.}}
 \scalebox{0.65}{
  \begin{tabular}[b]{lcccc}
   \toprule
    & Label & Worker Bias & Item Bias & Cognitive Bias \\
   \midrule
   \cite{dawid1979} & categorical & \checkmark & &  \\
   \midrule
   \cite{whitehill2009}, \cite{welinder2010}& categorical & \checkmark & \checkmark &  \\
   \midrule
   \cite{raykar2011} & ordinal & \checkmark &  &  \\
   \midrule
   \cite{baba2013},  \cite{nguyen2016} & ordinal & \checkmark & \checkmark &  \\
   \midrule
   \cite{zhuang2015kdd} & categorical / ordinal &  &  & \checkmark \\
   \midrule \midrule
   {\sc ImpCIM/CDM/ImpCDM} (proposed) & ordinal & \checkmark & \checkmark & \checkmark \\
   \bottomrule
  \end{tabular}}
 \label{method_overview}
\end{table*}

Finally,  to characterize the technical aspects of this study, we present a taxonomic comparison of the generative model structure in this study with related studies in crowdsourcing label aggregation in Table~\ref{method_overview}.
Whether the label format is categorical or ordinal, most of the existing models mainly focus only on ways to incorporate worker biases and item biases.
On the other hand, \cite{zhuang2015kdd} considered the cognitive bias in batch annotations and proposed a model in which the annotation strategy is switched by the biases, but they did not target the biases of workers or items themselves.
Our important technical contribution is the incorporation of the cognitive bias perspective into the framework with generative models that address the worker and item biases.

\section{Bias Investigation in Multi-Criteria Quality Evaluation}\label{dataset_section}
To verify the presence of inter-criteria cognitive biases in multi-criteria quality \textcolor{black}{assessments}, we construct text writing datasets that have multiple \textcolor{black}{evaluation} aspects, and then collect responses that rate them from each aspect. 
The statistical analysis reveals how inter-criteria cognitive biases affect multi-criteria assessment. 

\subsection{Text Writing Datasets}

\begin{table*}[t]
 \centering
 \caption{Writing task settings.}
  \begin{tabular}[b]{lll}
   \toprule
    & Review & Profile \\
   \midrule
   Task & Write a review about & Write workers' experiences \\
   description & a favorite restaurant & or skills \\
   \midrule
   \#published tasks & 50 & 50 \\
   \midrule
   Max. \#tasks allowed & 1 & 1 \\
   per worker &  &  \\
   \midrule
   Reward for each task &  \$1.36 &  \$1.00\\
   \bottomrule
  \end{tabular}
 \label{writing_task_setting}
\end{table*}

We start by creating a set of evaluation targets using crowdsourcing.
We use short text writings as quality evaluation targets, as they are relatively easy and allow for various aspects of evaluation. 
We set up two writing topics; the first topic is ``customer reviews about restaurants'' where a crowdworker is asked to introduce one of his/her favorite restaurants. 
The other topic is ``personal profile statements'' where we ask them to describe their experiences or skills assuming an application resume for a job in an advertising agency.
In the task guidelines, we advise crowd writers not to mention the actual restaurant names or personal information to ensure anonymity and privacy. 
We publish the text writing tasks in the Lancers\footnote{\url{https://lancers.jp}} crowdsourcing marketplace. 
Table~\ref{writing_task_setting} describes the setting of the writing tasks.

For the \textcolor{black}{review} and \textcolor{black}{profile} \textcolor{black}{tasks}, we \textcolor{black}{obtained} 50 reviews and 37 profiles written by different crowdworkers\textcolor{black}{,} respectively. 
We \textcolor{black}{reviewed} all of them before using them as targets for multi-criteria assessment and \textcolor{black}{discarded} invalid texts and useless information. 
The final text writing datasets \textcolor{black}{contained} 50 reviews and 36 profiles.

\subsection{Evaluation Datasets}

\begin{table*}[t]
 \centering
 \caption{Review evaluation task settings.}
  \begin{tabular}[b]{lll}
   \toprule
    & INDV & SIMUL \\
   \midrule
   Task & Evaluate a review & Evaluate a review \\
   description & in a criterion & in five criteria \\
   \midrule
   \#evaluation & 50 & 50 \\
   targets & & \\
   \midrule
   \#published & 25,000 & 5,000 \\
   tasks & & \\
   \midrule
   Max. \#tasks allowed & Unlimited & Unlimited \\
   per worker & & \\
   \midrule
   Reward & \$0.02 & \$0.05 \\
   for each task & & \\
   \bottomrule
  \end{tabular}
 \label{review_evaluation_task_setting}
\end{table*}

\begin{table*}[t]
 \centering
 \caption{Profile evaluation task settings.}
 \begin{tabular}[b]{lll}
   \toprule
    & INDV & SIMUL \\
   \midrule
   Task & Evaluate a profile & Evaluate a profile \\
   description & in a criterion & in five criteria \\
   \midrule
   \#evaluation & 36 & 36 \\
   targets & & \\
   \midrule
   \#published & 18,000 & 3,600 \\
   tasks & & \\
   \midrule
   Max. \#tasks allowed & Unlimited & Unlimited \\
   per worker & & \\
   \midrule
   Reward & \$0.02 & \$0.05 \\
   for each task & & \\
   \bottomrule
  \end{tabular}
 \label{profile_evaluation_task_setting}
\end{table*}

Next, we \textcolor{black}{created} evaluation datasets that \textcolor{black}{were} collections of the crowd evaluation results with respect to multiple criteria (which will be described later.)
By using these datasets, we \textcolor{black}{validated} the existence of inter-criteria cognitive biases in \textcolor{black}{the} multi-criteria assessment. 
To this end, we \textcolor{black}{considered} a comparative analysis based on two types of evaluation settings: individual and simultaneous assessment.
In the individual assessment, a worker rates each text with only one of all criteria; in other words, when judging based on a single criterion, the worker does not consider the others.
In contrast, in the simultaneous assessment, a worker assesses each text in all the criteria at the same time; thus the worker assesses each criterion while noticing the others.
We \textcolor{black}{treated} the former as a dataset without multi-criteria assessment biases, and the latter as the biased one. 
By comparing these two settings, we \textcolor{black}{analyzed} how simultaneous assessment \textcolor{black}{affected} the worker responses. 
\textcolor{black}{Unfortunately}, we \textcolor{black}{could not} completely remove such biases by the individual assessment, because workers can judge each independent criterion based on the global impression of the target consciously or unconsciously, or they might suffer from effects of other evaluation tasks in which they participated before; however, it \textcolor{black}{seemed} reasonable to assume that responses collected under the individual setting \textcolor{black}{were} less susceptible to inter-criteria cognitive biases than the simultaneous setting.

Previous studies utilized criteria for text assessment in terms of i) grammatical writing style, ii) written information, and iii) connectivity of texts~\citep{Lester1997,Mellish1998,Vougiouklis2018NeuralWikipedian}. 
Following their approaches, we \textcolor{black}{picked} four criteria: 
\begin{itemize}
    \item \emph{coherence} (from (iii)) evaluating whether the content is logically connected,
    \item \emph{organization} (from (ii)) evaluating whether the information is organized and concise, 
    \item \emph{writing style} (from (i)) evaluating whether the text is correctly written in terms of grammar and notation (e.g., spelling errors and omissions), and
    \item \emph{readability} (from (iii)) evaluating whether the text is easy to read.
\end{itemize}
\textcolor{black}{We included} one additional criterion called \emph{overall} for the comprehensive evaluation of text quality.
On each criterion, workers \textcolor{black}{judged} the quality of texts using a five-point Likert scale, where higher rating scores correspond to higher quality. 

We \textcolor{black}{published} the evaluation tasks of our text writing datasets on Lancers. 
For each text, our study \textcolor{black}{required} six evaluation tasks: one for each of the five criteria in the individual setting and the other one for the simultaneous assessment of all five criteria.
Each question in the evaluation tasks \textcolor{black}{consisted} of a target text, the definitions of the criteria, and five check boxes for a five-point Likert scale for each criterion. 
Each of the individual assessments tasks \textcolor{black}{used} only one criterion per query, while the simultaneous assessment task \textcolor{black}{used} all five criteria.
The questions \textcolor{black}{were} provided in a random order, and the criteria in each question \textcolor{black}{were} also shuffled in simultaneous assessments in order to avoid positional biases.
In the remainder of this paper, we denote the individual condition by ``INDV'' and the the simultaneous condition by ``SIMUL''. 
Table~\ref{review_evaluation_task_setting} and~\ref{profile_evaluation_task_setting} describe the settings of the evaluation tasks.

Since we \textcolor{black}{published} multiple evaluation tasks in different settings in a relatively short time period, the collected responses possibly \textcolor{black}{introduced} inter-task biases~\citep{cai2016,aipe2018}.
Specifically, in INDV tasks, where only one criterion \textcolor{black}{was} provided, if workers repeatedly \textcolor{black}{participated in} different tasks, they could unconsciously make evaluations that \textcolor{black}{were} influenced by other criteria that they already evaluated in previous tasks. 
To avoid such situations as much as possible, we \textcolor{black}{published} each task one by one on a different day. 
Moreover, to prevent workers from joining the INDV tasks after having evaluated all criteria in SIMUL tasks, we \textcolor{black}{published} the SIMUL tasks after all the INDV tasks \textcolor{black}{were} finished. 

After all the evaluation tasks \textcolor{black}{were} completed, for each target-criterion pair, we \textcolor{black}{obtained} 100 INDV and 74 SIMUL ratings from 549 workers for restaurant reviews, and 100 INDV and 100 SIMUL ratings for profile statements from 589 workers. 
In the review assessments, we \textcolor{black}{observed} that 169 workers participated in different individual assessment tasks for different criteria.
Additionally, 137 workers completed both INDV and SIMUL assessments, and particularly, 19 of those assessed all target-criterion pairs in both assessment conditions.
Similary, in the profile assessments, we \textcolor{black}{observed} 178 workers participating in different individual tasks, 262 workers completing both INDV and SIMUL assessments, and 23 of those assessing all target-criterion pairs in both assessment conditions.
Such repeated participation of the same workers in different tasks is not desirable, as we want to compare cases where multiple criteria are assessed simultaneously with cases where they are independently assessed. 
In order to reduce such bias, we \textcolor{black}{released} each task for INDV on a different day, after all SIMUL tasks \textcolor{black}{were} completed. 
Thus, we determined that this overlap would not seriously affect the analysis. 

\subsection{Bias Analysis}
We \textcolor{black}{then} \textcolor{black}{analyzed} the collected response data to investigate how inter-criteria cognitive biases \textcolor{black}{affected} crowdworker ratings. 
Research in psychology has focused on two types of rater biases: the \textit{leniency} error and \textit{halo} error~\citep{hoyt2000}. 
While leniency explains the tendency of rater responses across targets, halo error is conceptualized as the raters' failure to discriminate between multiple criteria~\citep{saal1980}, which affects the inter-criteria mean and variance of the ratings, due to the prominence of criteria or raters' impressions of the target~\citep{balzer1992}. 
Following the literature, we \textcolor{black}{investigated} the inter-criteria effect in our datasets in terms of the mean and variance of ratings.

\begin{figure}[t]
\centering
 \begin{minipage}{0.49\hsize}
 \begin{center}
  \includegraphics[width=4.1cm]{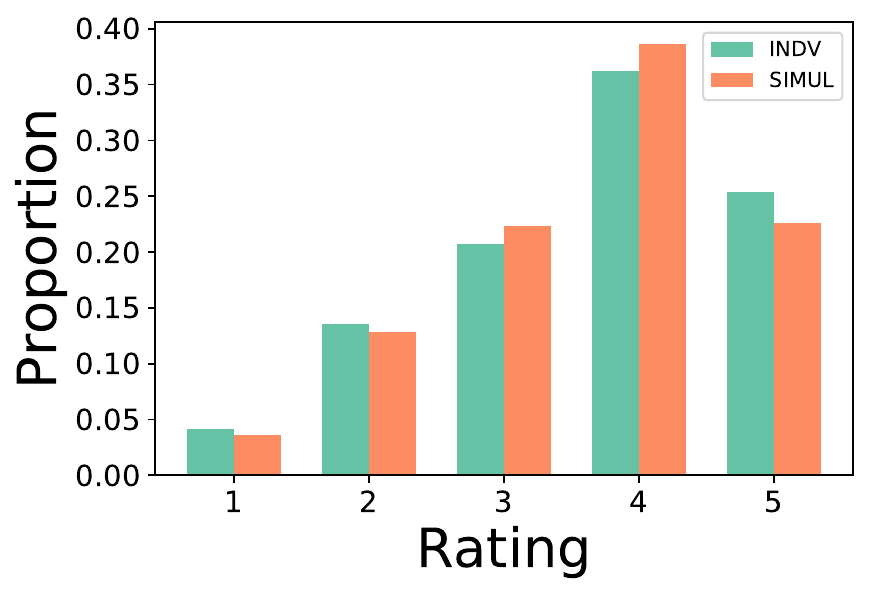}
  \subcaption{Review}
  \label{review_ratings}
 \end{center}
 \end{minipage}
 \begin{minipage}{0.49\hsize}
  \begin{center}
   \includegraphics[width=4.1cm]{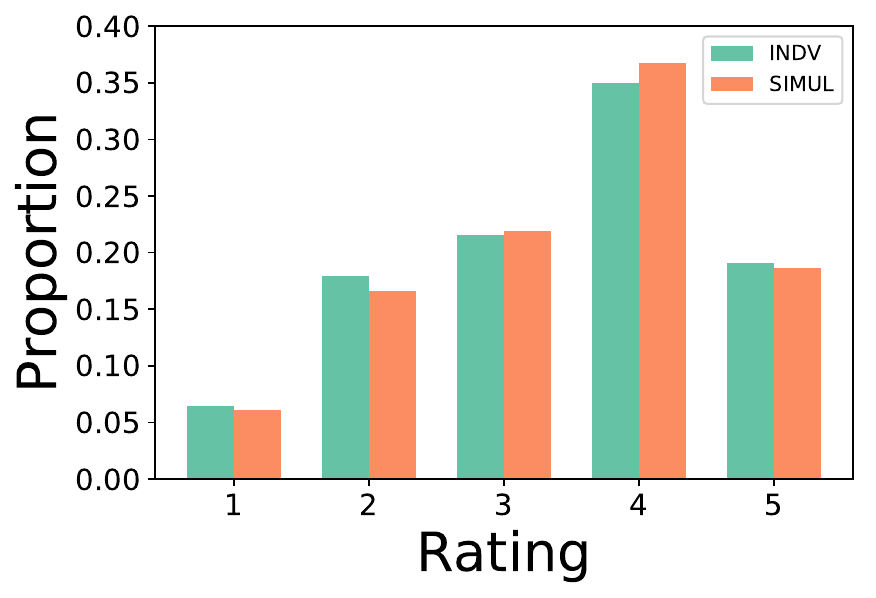}
   \subcaption{Profile}
   \label{profile_ratings}
  \end{center}
 \end{minipage}
 \caption{Distributions of worker responses of (a) review dataset and (b) profile dataset. 
 The left bar of each rating corresponds to INDV and the right bar to SIMUL.}
 \label{rating_distributions}
\end{figure} 

We first \textcolor{black}{examined} the tendency of the five graded ratings in the dataset. 
Figure \ref{rating_distributions} shows the proportion of each grade in the total response data for each assessment type.
Workers generally \textcolor{black}{hesitated} to give extreme ratings such as 1 and 5, and this trend \textcolor{black}{became} stronger in SIMUL given that occurrences of 3 and 4 \textcolor{black}{were} more frequent in SIMUL than in INDV. 
This concentration of responses owing to the simultaneous evaluation of multiple criteria \textcolor{black}{reduced} the variance of the assessment scores, and hence likely \textcolor{black}{suggested} the existence of a halo effect. 

\begin{figure}[t]
\centering
 \begin{minipage}{0.49\hsize}
 \begin{center}
  \includegraphics[width=4.1cm]{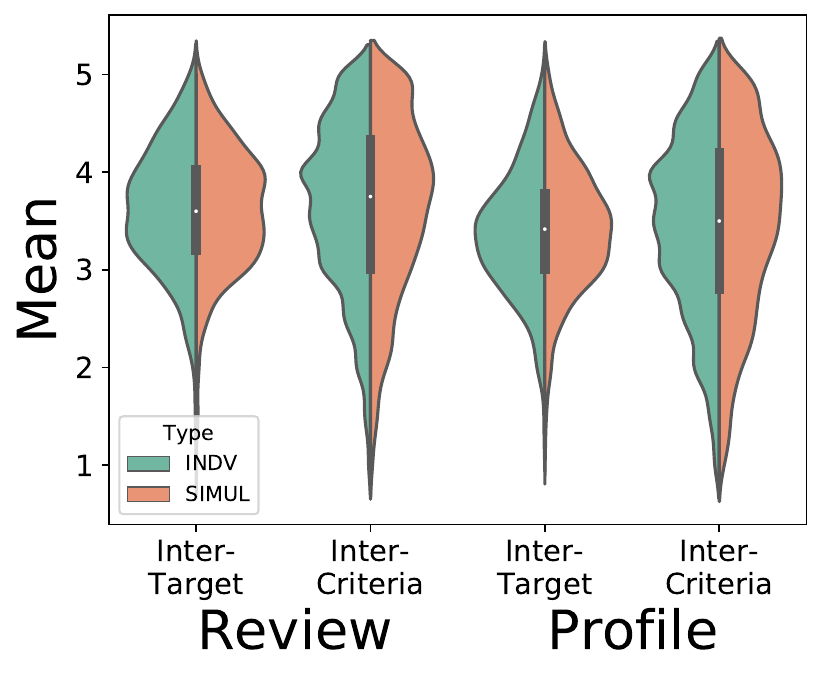}
  \subcaption{Distributions of mean}
  \label{mean_violin}
 \end{center}
 \end{minipage}
 \begin{minipage}{0.49\hsize}
  \begin{center}
   \includegraphics[width=4.1cm]{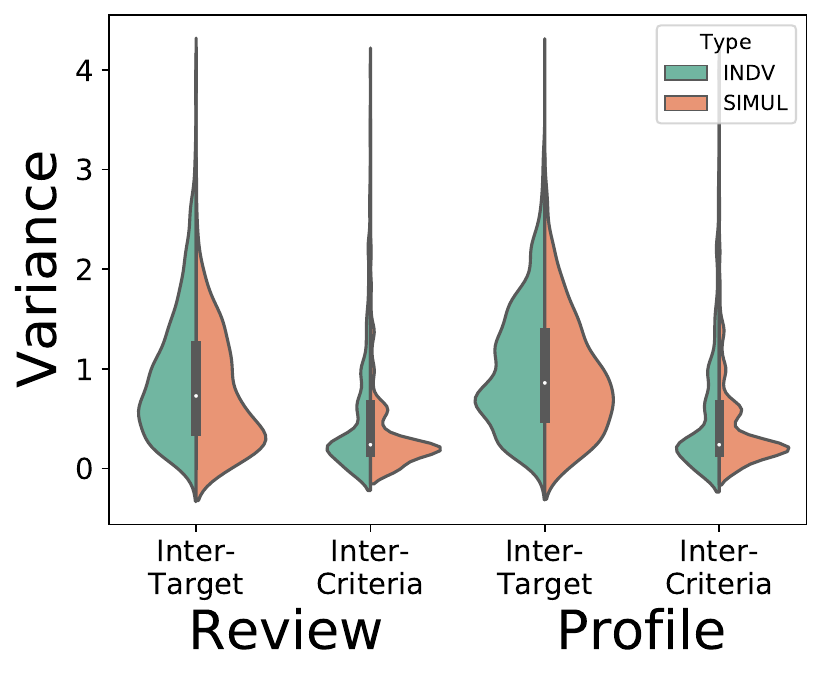}
   \subcaption{Distributions of variance}
   \label{variance_violin}
  \end{center}
 \end{minipage}
 \caption{Violin plots of distributions of inter-target or inter-criteria (a) mean and (b) variance of worker responses. The left area of each violin corresponds to INDV and the right area to SIMUL.}
 \label{mean_variance_violin}
\end{figure}

\begin{table}[t]
 \centering
 \caption{Mean and standard deviation (SD) of the (a) mean and (b) variance of worker responses.  ``Targets" and ``Criteria" in the columns stand for inter-target and inter-criteria, respectively.}
 \begin{center}
  \subcaption{Distributions of mean}
  \begin{tabular}[b]{lcccc}
   \toprule & \multicolumn{2}{c}{Review} & \multicolumn{2}{c}{Profile} \\
   \cmidrule{2-3} \cmidrule{4-5}
    & Targets & Criteria & Targets & Criteria \\
   \midrule
   Mean (INDV) & 3.624 & 3.643 & 3.381 & 3.428 \\
   Mean (SIMUL) & 3.615 & 3.638 & 3.426 & 3.452 \\
   SD (INDV) & 0.624 & 0.850 & 0.629 & 0.900 \\
   SD (SIMUL) & 0.619 & 0.905 & 0.637 & 0.957 \\
   \bottomrule
   \label{mean_distribution_statistics}
  \end{tabular}
 \end{center}
 \begin{center}
  \subcaption{Distributions of variance}
  \begin{tabular}[b]{lcccc}
   \toprule & \multicolumn{2}{c}{Review} & \multicolumn{2}{c}{Profile} \\
   \cmidrule{2-3} \cmidrule{4-5}
    & Targets & Criteria & Targets & Criteria \\
   \midrule
   Mean (INDV) & 0.947 & 0.538 & 1.019 & 0.558 \\
   Mean (SIMUL) & 0.744 & 0.338 & 0.907 & 0.404 \\
   SD (INDV) & 0.652 & 0.616 & 0.629 & 0.619 \\
   SD (SIMUL) & 0.573 & 0.395 & 0.581 & 0.423 \\
   \bottomrule
   \label{variance_distribution_statistics}
  \end{tabular}
 \end{center}
 \label{mean_variance_statistics}
\end{table}

To further investigate the inter-criteria cognitive biases, we \textcolor{black}{inspected} the inter-criteria mean and variance of ratings over criteria. 
When computing the inter-criteria statistics, we \textcolor{black}{discarded} target-worker pairs that produced only one answer in the evaluation procedure, so that we only \textcolor{black}{considered} the effect over multiple criteria. 
For comparison, we also \textcolor{black}{checked} the inter-target statistics for each worker-criterion pair, while discarding worker-criterion pairs with only one answer. 
Figure~\ref{mean_violin} and Table~\ref{mean_distribution_statistics} show that the inter-target and inter-criteria means are both similarly distributed in INDV and SIMUL. 
\textcolor{black}{Alternatively}, we \textcolor{black}{observed} a significant difference in the variance of the inter-criteria mean distribution between INDV and SIMUL in the review dataset
[inter-target: $F(1,1574)=0.045, p = 0.831$; inter-criteria: $F(1,9113)=17.5, p < .001$] 
and the profile dataset 
[inter-target: $F(1,1821)=0.165, p = 0.684$; inter-criteria: $F(1,7609)=14.604, p < .001$]
using the two-sided F-test.
Therefore, in SIMUL, there could be cases where workers over or under-\textcolor{black}{estimated} the criteria affected by a portion of the criteria with significant quality. 
On the other hand, Figure~\ref{variance_violin} and Table~\ref{variance_distribution_statistics} show the inter-target and inter-criteria variances.
For all the four  conditions, we \textcolor{black}{observed} a significant difference between INDV and SIMUL 
[inter-target of review: $t(1573)=6.466, p<.001$;
inter-criteria of review: $t(9112)=18.885, p<.001$;
inter-target of profile: $t(1820)=2.948, p<.001$;
inter-criteria of profile: $t(7608)=12.776, p<.001$]
using the two-sided Welch's $t$-test.
\textcolor{black}{Since} the inter-target and inter-criteria variances seemingly do not follow normal distributions as shown in Figure~\ref{variance_violin},
we also \textcolor{black}{applied} the  non-parametric Brunner-Munzel test~\citep{brunner2000}. Consequently, we also \textcolor{black}{observed} a significant difference between INDV and SIMUL for all four conditions
[inter-target of review: $W(1575) = -6.08, p < .001$;
inter-criteria of review: $W(9114) = -17.65, p < .001$;
inter-target of profile: $W(1822) = -3.53, p < .001$;
inter-criteria of profile: $W(7610) = -10.46, p < .001$].
These results \textcolor{black}{confirmed} the existence of a halo error, which \textcolor{black}{reduced} the ratings variance over criteria, as well as \textcolor{black}{reduced} the inter-target variance in our response datasets.

In summary, we \textcolor{black}{collected} multi-criteria assessment datasets on two text writing tasks using crowdsourcing and \textcolor{black}{analyzed} the effect of inter-criteria cognitive biases.
Our statistical investigation \textcolor{black}{revealed} that simultaneous evaluation of multiple criteria causes over- or under-estimation with lower variance than their independent evaluation.
This result leads us to the question of how such biases should be addressed.
In the following sections, we consider the issue of bias mitigation in opinion aggregation models. 

\section{Bias Mitigation in Multi-Criteria Quality Evaluation}\label{proposedmethod_section}
In this section, we propose structures for Bayesian opinion aggregation models to mitigate the effect of inter-criteria cognitive biases.
Based on the previous observation, we split the cognitive effect into two components, the mean and the variance, of worker responses, and propose two structures corresponding to them respectively. 
We first present a basic aggregation model for multi-criteria assessment, which is a na\"ive extension of the single-criterion model. 
We further propose two model structures to incorporate inter-criteria biases into the base model. 

\subsection{Problem Setting}
We treat bias mitigation as a rating aggregation problem to estimate the quality of the target items by modeling the generative process of a set of worker ratings. 
\textcolor{black}{Assuming} we have a collection of $I$ evaluation targets (i.e., the texts in our datasets)\textcolor{black}{,} $M$ evaluation criteria, and $J$ workers assess the quality of each target in terms of each criterion on a five-point Likert scale.
The response from worker $j$ to criterion $m$ of target $i$ is denoted by $x_{ij}^{(m)} \in \{ 1, 2, \dots, 5 \}$.
Given a set of five-graded responses $\{ x_{ij}^{(m)} \}_{i,j,m}$, rating aggregation models infer the potential quality $t_i$ of target $i$ and criteria quality $t_i + q_{i}^{(m)}$ in each criterion $m$, where $q_i^{(m)}$ represents the difference in the quality of the $m$-th criterion from the potential quality.

\subsection{Base Model: Criteria Independent Model ({\sc CIM})}

\begin{figure}[tb]
 \begin{center}
  \includegraphics[width=8cm]{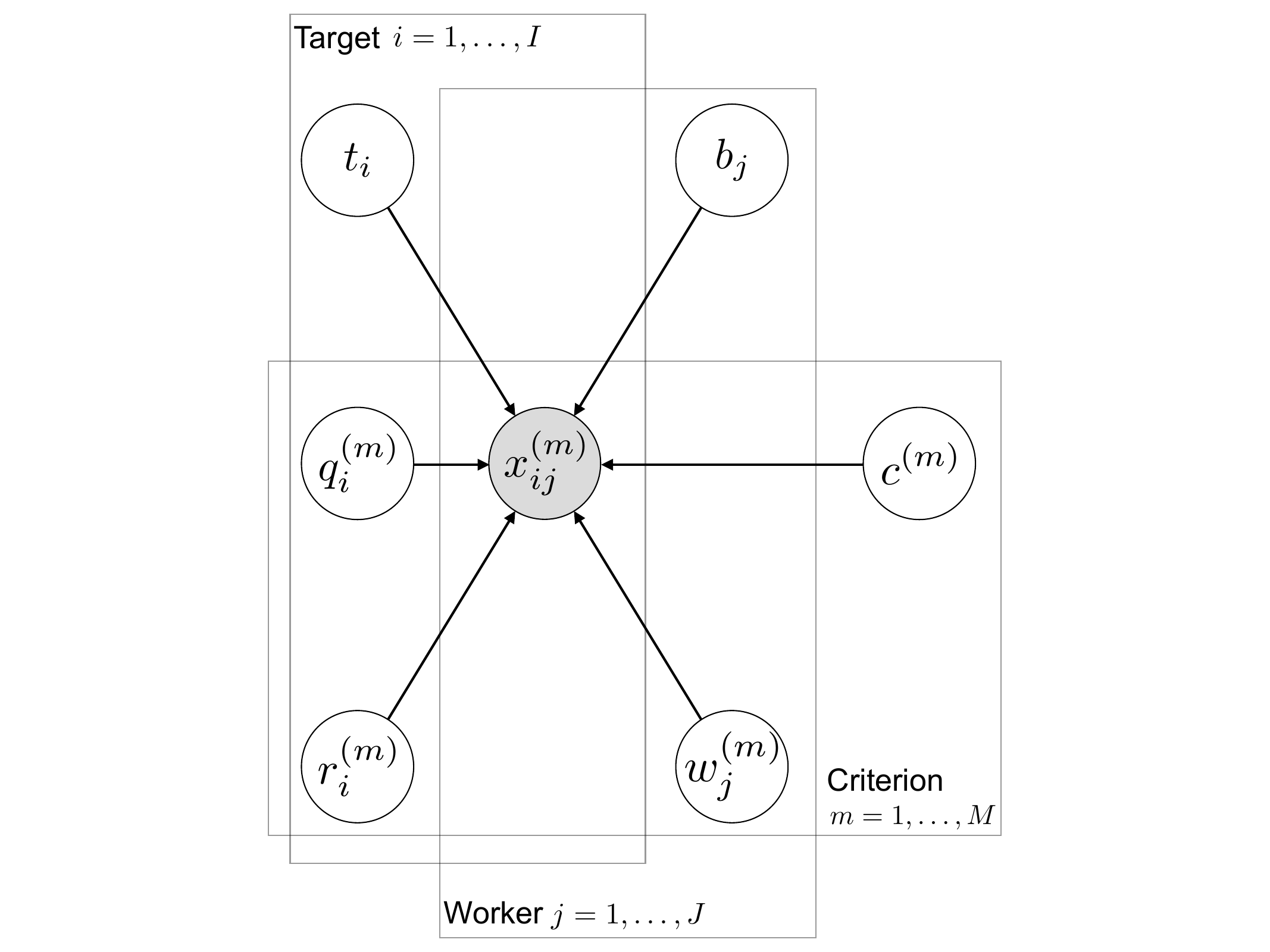}
  \caption{Graphical model of {\sc CIM}. Observation $x_{ij}^{(m)}$ is indicated by a shaded circle.}
  \label{graphical_models_CIM}
 \end{center}
\end{figure}

\begin{figure}[tb]
 \begin{center}
  \includegraphics[width=8cm]{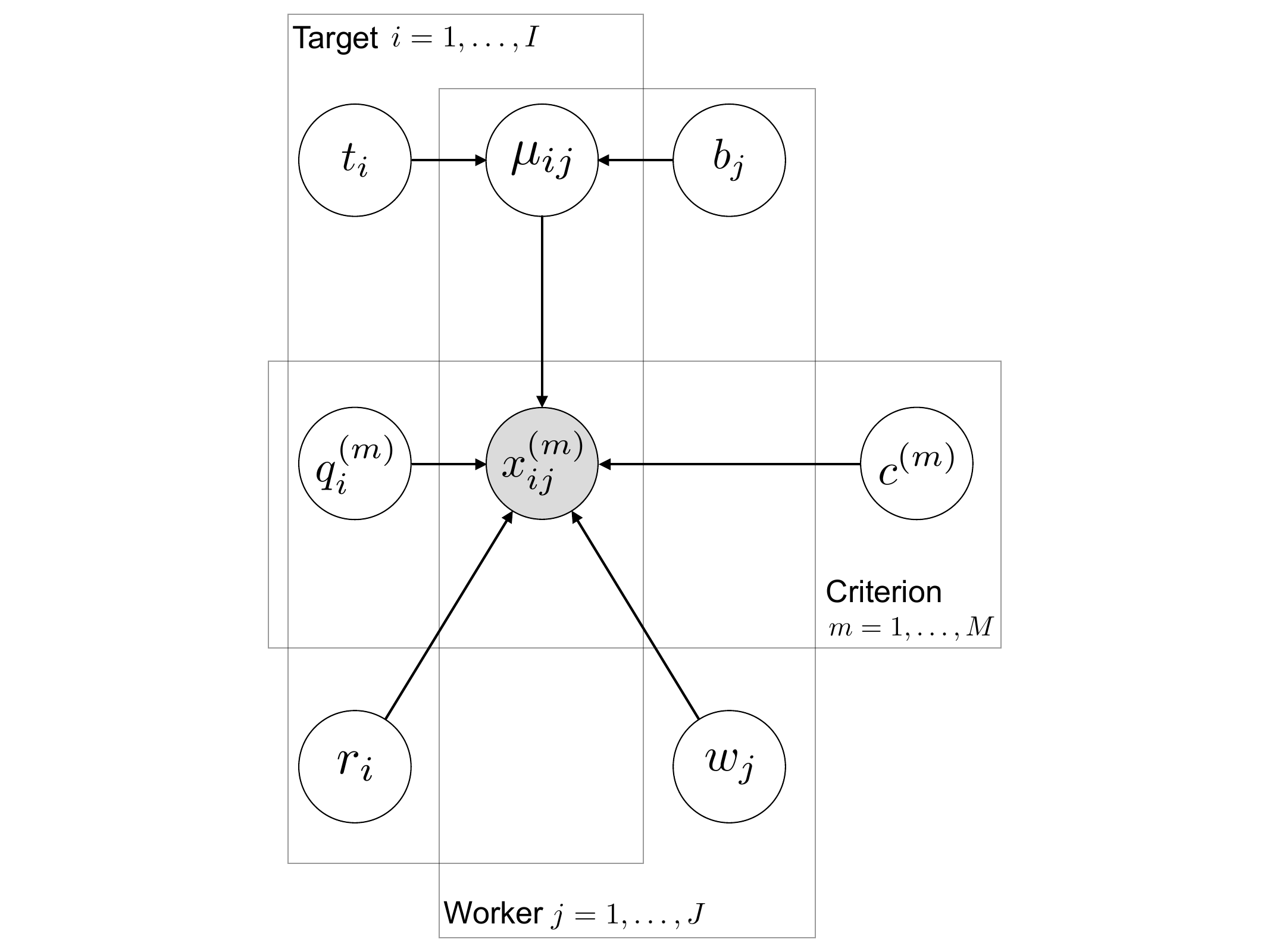}
  \caption{Graphical model of {\sc ImpCDM}. Observation $x_{ij}^{(m)}$ is indicated by a shaded circle.}
  \label{graphical_models_impCDM}
 \end{center}
\end{figure}

Since our goal is to design a rating aggregation model that accounts for cognitive biases, we begin by building a na\"ive rating aggregation model for multi-criteria assessment as our base model.
Although several aggregation models have been proposed for ordinal ratings~\citep{raykar2010,baba2013}, the focus of their research is the single-criterion assessment.
To evaluate our proposed bias mitigation structures, instead of directly employing their models as base models, we construct a base model using structures and parameters introduced in the existing model.

As shown in Table~\ref{method_overview}, one core concept of generative aggregation models is the bias of individual workers and items.
Most of related studies assume that observed labels depend on the combination of true labels, worker biases (e.g., their ability or expertise), and item biases (e.g., feature variations or difficulty of annotation).
Categorical labels are modeled using the probability of being given correct annotations~\citep{dawid1979,whitehill2009,welinder2010}, and the generative process of ordinal labels assumes that observed values are generated  depending on latent scores~\citep{baba2013,nguyen2016}.
Since our focus is close to the latter, we design our model based on the generative distribution for ordinal labels. 

The latent scores for ordinal labels depend on the discrepancy between the worker's perceived quality and the true quality.
One way to define the latent score is to sum the worker bias and item bias to the true quality~\citep{baba2013}.
Another way is to introduce item features and  and use their weighted sum~\citep{nguyen2016}.
Because it is not always possible to extract features from the data generated by the crowd workers, we adopt the former approach as the structure of our latent scores for the sake of generality of the proposed method.

There are two typical ways to model the generative process of ordinal ratings: the normal distribution and the graded response model. 
The normal distribution is not an inherently proper model, but is often used as an approximation in practice, such as in recommendation systems~\citep{salakhutdinov2008}.
The graded response model is a standard model of graded responses used in the item response theory~\citep{samejima1969}. 
In this study, we employ normal distribution for simplicity because it can easily model the mean and variance of worker responses.

Figure~\ref{graphical_models_CIM} shows the simple aggregation model that we call the criteria independent model ({\sc CIM}).
The {\sc CIM} models the generative process of $x_{ij}^{(m)}$ using the normal distribution.
As discussed above,
we set the mean of the distribution as the sum of the quality of the $m$-th criterion $t_i + q_i^{(m)}$, the bias of the $j$-th worker $b_j$, and the bias of the $m$-th criterion $c^{(m)}$, where we attach the criterion bias to express the difference in rating tendency among different criteria.
As the variance of the normal distribution, we take the sum of the target-criterion variance $r_i^{(m)}$ and worker-criterion variance $w_j^{(m)}$ to express the different rating variability among targets and workers in each criterion.
In short,
\textcolor{black}{the} {\sc CIM} generates the graded response $x_{ij}^{(m)} \in \{ 1, 2, \dots, 5 \}$ as follows:
\begin{eqnarray}
\label{CIM_x}
x_{ij}^{(m)} \sim {\mathcal N} (t_i + q_i^{(m)} + b_j + c^{(m)}, r_i^{(m)} + w_j^{(m)}).
\end{eqnarray}
The worker bias in \textcolor{black}{the} {\sc CIM} represents the tendency of workers who always over- or underestimate targets, and thus, this bias does not consider the inter-criteria effect.
Similarly, the other parameters also do not consider the inter-criteria relations.
In our models, we set normal distribution priors for $t_i$, $q_i^{(m)}$, $b_j$, and $c^{(m)}$, and gamma distribution priors for $r_i^{(m)}$ and $w_j^{(m)}$.

\subsection{Proposed Model: Impression Criteria Dependent Model ({\sc ImpCDM})}
We extend the previous base model {\sc CIM} in two ways to consider the inter-criteria cognitive biases. 
Our first proposed structure considers the bias in terms of mean, where workers tend to evaluate each criterion based on a global impression of the target.
This effect causes over or under-estimation over all criteria of the target, which causes the mean ratings to tend toward higher or lower values. 
We model this phenomenon by introducing the impression parameter $\mu_{ij}$,
where we assume that impression of the $j$-th worker about the $i$-th target depends on either the potential quality of the target or the inherent bias of the worker; that is, targets with high potential quality tend to make a good impression on raters, and similarly, workers with a large positive inherent bias tend to feel good about the target.
Based on this assumption, the extended {\sc CIM} with the impression parameter generates five-graded responses as follows:
\begin{eqnarray}
\label{impCIM_x}
x_{ij}^{(m)} \sim {\mathcal N} (\mu_{ij}+ q_i^{(m)} + c^{(m)}, r_i^{(m)} + w_j^{(m)}),
\end{eqnarray}
where $\mu_{ij}$ is generated by the normal distribution with the potential quality $t_i$ and worker bias $b_j$:
\begin{eqnarray}
\label{gen_mu}
\mu_{ij} \sim {\mathcal N} (t_i + b_j, 1).
\end{eqnarray}
Through the additional sampling process from the true potential quality and the worker bias, we represent the variability of the impression from the true quality of the target.

The second proposed structure considers the inter-criteria cognitive bias in terms of variance, where worker ratings are likely to be similar across different criteria.
In the simultaneous assessment, a worker considers all the criteria of a target at once.
Based on this evaluation process, we assume that
the variance of responses is common among criteria in a task in which a worker assesses all the criteria of a target; in other words,
the target has its own variability or difficulty of evaluation among its criteria, and similarly, the worker has its own variability of evaluation among evaluation criteria.
This is formulated by simply removing the dependence on $m$ from the two variance parameters, $r_i^{(m)}$ and $w_j^{(m)}$ in~\eqref{CIM_x}, as follows:
\begin{eqnarray}
\label{CDM_x}
x_{ij}^{(m)} \sim {\mathcal N} (t_i + q_i^{(m)} + b_j + c^{(m)}, r_i + w_j).
\end{eqnarray}

Our final model imports the previous two proposed structures into the base model {\sc CIM}.
The model considers both the impression parameter and the common variance among criteria as follows:
\begin{eqnarray}
\label{impCDM_x}
x_{ij}^{(m)} \sim {\mathcal N} (\mu_{ij}+ q_i^{(m)} + c^{(m)}, r_i + w_j),
\end{eqnarray}
where $\mu_{ij}$ is generated by Eq.~\eqref{gen_mu}.
We refer to model~\eqref{impCDM_x} as the {\sc ImpCDM}. Its graphical model is shown in Figure~\ref{graphical_models_impCDM}.

\begin{table*}[t]
 \centering
 \caption{\color{black}Results on review evaluation datasets. Experimental results are shown in the mean of the correlation scores over $20$ trials. 
 The scores are in bold when the mean is higher than the other. The ``Potential'' \textcolor{black}{column} shows the results on the estimated $\{t_i\}_i$, and the other five columns correspond to the five evaluation criteria and the results on the estimated $\{ t_i + q_i^{(m)} \}_{i, m=1, 2, \dots, 5} $, respectively.\color{black}}
 \scalebox{0.78}{
 \begin{tabular}[t]{lcccccc}
  \toprule[0.1em] Model & Potential & Coherence & Organization & Writing style & Readability & Overall \\
  \midrule[0.08em]
{\sc CIM} (n=5) & 0.338 & 0.102 & 0.089 & 0.516 & 0.502 & 0.253 \\
{\sc CDM} (n=5) & 0.106 & 0.120 & 0.063 & \textbf{0.594} & 0.522 & 0.259 \\
{\sc ImpCIM} (n=5) & 0.414 & 0.181 & 0.184 & 0.486 & \textbf{0.566} & 0.339 \\
{\sc ImpCDM} (n=5) & \textbf{0.423} & \textbf{0.350} & \textbf{0.200} & 0.332 & 0.485 & \textbf{0.380} \\
\midrule[0.08em]
{\sc CIM} (n=6) & 0.359 & 0.105 & 0.278 & 0.559 & 0.523 & 0.158 \\
{\sc CDM} (n=6) & 0.084 & 0.230 & 0.168 & \textbf{0.776} & 0.597 & 0.280 \\
{\sc ImpCIM} (n=6) & 0.606 & 0.390 & \textbf{0.409} & 0.727 & \textbf{0.660} & 0.476 \\
{\sc ImpCDM} (n=6) & \textbf{0.687} & \textbf{0.500} & 0.366 & 0.624 & 0.631 & \textbf{0.519} \\
\midrule[0.08em]
{\sc CIM} (n=7) & 0.641 & 0.541 & 0.308 & \textbf{0.666} & 0.693 & 0.345 \\
{\sc CDM} (n=7) & 0.257 & 0.431 & 0.217 & 0.578 & 0.497 & 0.343 \\
{\sc ImpCIM} (n=7) & 0.675 & \textbf{0.641} & \textbf{0.433} & 0.649 & \textbf{0.700} & \textbf{0.603} \\
{\sc ImpCDM} (n=7) & \textbf{0.722} & 0.374 & 0.217 & 0.489 & 0.482 & 0.276 \\
\midrule[0.08em]
{\sc CIM} (n=8) & 0.440 & 0.125 & 0.265 & 0.631 & 0.683 & 0.393 \\
{\sc CDM} (n=8) & 0.110 & 0.393 & 0.337 & \textbf{0.713} & 0.662 & 0.574 \\
{\sc ImpCIM} (n=8) & 0.663 & 0.502 & \textbf{0.530} & 0.670 & 0.652 & \textbf{0.694} \\
{\sc ImpCDM} (n=8) & \textbf{0.733} & \textbf{0.515} & 0.469 & 0.627 & \textbf{0.702} & 0.671 \\
\midrule[0.08em]
{\sc CIM} (n=9) & 0.444 & 0.340 & 0.188 & 0.609 & 0.618 & 0.366 \\
{\sc CDM} (n=9) & 0.086 & 0.473 & 0.322 & \textbf{0.627} & 0.638 & 0.506 \\
{\sc ImpCIM} (n=9) & 0.571 & 0.591 & \textbf{0.420} & 0.560 & \textbf{0.699} & \textbf{0.675} \\
{\sc ImpCDM} (n=9) & \textbf{0.659} & \textbf{0.599} & 0.394 & 0.577 & 0.697 & 0.636 \\
\midrule[0.08em]
{\sc CIM} (n=10) & 0.314 & 0.078 & 0.135 & 0.432 & 0.508 & 0.275 \\
{\sc CDM} (n=10) & 0.037 & 0.204 & 0.086 & 0.632 & 0.535 & 0.286 \\
{\sc ImpCIM} (n=10) & 0.537 & 0.393 & \textbf{0.352} & \textbf{0.668} & \textbf{0.713} & \textbf{0.391} \\
{\sc ImpCDM} (n=10) & \textbf{0.602} & \textbf{0.523} & 0.246 & 0.619 & 0.633 & 0.330 \\
  \bottomrule[0.1em]
 \end{tabular}}
 \label{review_correlation_table}
\end{table*}

\begin{table*}[t]
 \centering
 \caption{\color{black}Results on profile evaluation datasets. Experimental results are shown in the mean of the correlation scores over $20$ trials. 
 The scores are in bold when the mean is higher than the other. The ``Potential'' \textcolor{black}{column} shows the results on the estimated $\{t_i\}_i$, and the other five columns correspond to the five evaluation criteria and the results on the estimated $\{ t_i + q_i^{(m)} \}_{i, m=1, 2, \dots, 5} $, respectively.\color{black}}
 \scalebox{0.78}{
 \begin{tabular}[t]{lcccccc}
  \toprule[0.1em] Model & Potential & Coherence & Organization & Writing style & Readability & Overall \\
  \midrule[0.08em]
{\sc CIM} (n=5) & 0.671 & \textbf{0.701} & 0.488 & \textbf{0.376} & \textbf{0.635} & 0.655 \\
{\sc CDM} (n=5) & 0.469 & 0.653 & 0.299 & 0.333 & 0.534 & 0.543 \\
{\sc ImpCIM} (n=5) & 0.650 & 0.689 & \textbf{0.509} & 0.238 & 0.546 & \textbf{0.702} \\
{\sc ImpCDM} (n=5) & \textbf{0.707} & 0.674 & 0.242 & 0.300 & 0.565 & 0.597 \\
\midrule[0.08em]
{\sc CIM} (n=6) & \textbf{0.734} & 0.674 & 0.554 & 0.641 & 0.366 & \textbf{0.609} \\
{\sc CDM} (n=6) & 0.650 & \textbf{0.705} & \textbf{0.576} & \textbf{0.666} & \textbf{0.560} & 0.584 \\
{\sc ImpCIM} (n=6) & 0.615 & 0.493 & 0.471 & 0.577 & 0.416 & 0.555 \\
{\sc ImpCDM} (n=6) & 0.672 & 0.339 & 0.389 & 0.250 & 0.422 & 0.609 \\
\midrule[0.08em]
{\sc CIM} (n=7) & 0.604 & 0.701 & 0.362 & 0.633 & 0.536 & 0.561 \\
{\sc CDM} (n=7) & 0.660 & 0.657 & 0.530 & 0.572 & \textbf{0.564} & 0.572 \\
{\sc ImpCIM} (n=7) & 0.621 & 0.668 & \textbf{0.535} & 0.603 & 0.532 & 0.607 \\
{\sc ImpCDM} (n=7) & \textbf{0.669} & \textbf{0.714} & 0.515 & \textbf{0.650} & 0.534 & \textbf{0.617} \\
\midrule[0.08em]
{\sc CIM} (n=8) & 0.730 & 0.743 & 0.695 & 0.549 & 0.542 & 0.759 \\
{\sc CDM} (n=8) & 0.380 & 0.556 & 0.544 & 0.317 & 0.452 & 0.730 \\
{\sc ImpCIM} (n=8) & 0.732 & \textbf{0.748} & \textbf{0.767} & 0.636 & \textbf{0.606} & \textbf{0.773} \\
{\sc ImpCDM} (n=8) & \textbf{0.803} & 0.629 & 0.729 & \textbf{0.643} & 0.573 & 0.523 \\
\midrule[0.08em]
{\sc CIM} (n=9) & 0.758 & 0.752 & 0.616 & 0.695 & 0.662 & 0.681 \\
{\sc CDM} (n=9) & 0.717 & 0.804 & \textbf{0.662} & \textbf{0.774} & \textbf{0.669} & 0.679 \\
{\sc ImpCIM} (n=9) & 0.717 & \textbf{0.815} & 0.650 & 0.624 & 0.637 & \textbf{0.696} \\
{\sc ImpCDM} (n=9) & \textbf{0.765} & 0.661 & 0.552 & 0.520 & 0.555 & 0.669 \\
\midrule[0.08em]
{\sc CIM} (n=10) & 0.736 & 0.685 & 0.545 & 0.461 & 0.548 & 0.582 \\
{\sc CDM} (n=10) & 0.816 & 0.805 & 0.673 & 0.575 & 0.518 & 0.618 \\
{\sc ImpCIM} (n=10) & 0.716 & \textbf{0.814} & 0.701 & \textbf{0.601} & 0.520 & \textbf{0.631} \\
{\sc ImpCDM} (n=10) & \textbf{0.822} & 0.644 & \textbf{0.747} & 0.458 & \textbf{0.549} & 0.382 \\
\bottomrule[0.1em]
 \end{tabular}}
 \label{profile_correlation_table}
\end{table*}
\section{Experiments}\label{experiments_section}
We conduct experiments using our response datasets to verify that the rating aggregation models using our proposed structures successfully mitigate the inter-criteria cognitive biases.

\subsection{Evaluation Metric}
We measure the bias mitigation performance of our rating aggregation models based on the prediction accuracy of the potential quality of the criteria. 
Measuring the prediction accuracy of aggregation models requires ground truths, i.e., the true quality of the texts in our dataset;
however, the true quality of the evaluation targets (reviews and profiles) is unknown as they are created by crowdworkers. 

The previous study by \cite{snow2008} has demonstrated that averaging a sufficient number of non-expert annotations achieves comparable quality to expert annotation in several tasks.
Based on their observation, we define the true quality as the mean of all (in our case, 100)  responses obtained under the individual assessment condition.
We estimate the potential and criteria quality using SIMUL responses, and compute the correlation coefficient $\rho$ between the estimated quality and the ``ground truth" quality.
In other words, we assume the average of 100 INDV worker responses as the ground truth for each (target, criterion)-pair, and try to estimate it from a limited number (5 to 10) of SIMUL responses. 

Note that, we use the Spearman's ranking correlation as the performance metric in our experiments, but other similar performance metrics gave no significant quantitative differences. 

\subsection{Experimental Setup}
To infer our Bayesian aggregation models, we utilize the maximum a posteriori estimation using the Adam optimizer to estimate the joint probability of the parameters.
We set the prior of $t_i$ to be the normal distribution centered at the center of the five-point scale; i.e.,
\begin{eqnarray}
t_i \sim {\mathcal N} (3, 1),
\end{eqnarray}
and the priors of $q_i^{(m)}, b_j$, and $c^{(m)}$ are set to be the normal distribution with zero mean and a variance of 1, as follows:
\begin{eqnarray}
q_i^{(m)}, b_j, c^{(m)} \sim {\mathcal N} (0, 1).
\end{eqnarray}
To give the priors for variance parameters $r_i^{(m)}, w_j^{(m)}, r_i$, and $w_j$, which are positive parameters, we assume the gamma distribution with shape $\alpha=2$ and rate $\beta=2$ as follows:
\begin{eqnarray}
r_i^{(m)}, w_j^{(m)}, r_i, w_j \sim {\rm Gamma} (2, 2).
\end{eqnarray}
Due to the \textcolor{black}{identification} difficulty \textcolor{black}{of} the optimal initialization, we run the inference $20$ times using different initial values and compute the correlation each time.

The size of the datasets is a key factor for the prediction performance of \textcolor{black}{the} aggregation models; thus, we select a subset of workers and compare the results with \textcolor{black}{a} different numbers of workers.
In order to make the situation similar to the actual use of crowdsourcing, where only a limited number of workers are employed, we randomly sample five to ten workers from those who evaluated all reviews or profiles in the simultaneous assessments.
The total numbers of workers who can be assigned to the estimation are $19$ in the review and $23$ in the profile \textcolor{black}{datasets}, respectively, and we resample workers in each of the $20$ estimation iterations.

\color{black}
Furthermore, in order to validate how each proposed model structure contributes to bias mitigation, we also prepare subset models of {\sc ImpCDM} and measure their performance.
The first subset model has the proposed mean structure of {\sc ImpCDM} and the \textcolor{black}{same} variance structure as {\sc CIM}, which we call {\sc ImpCIM}.
The second subset model has the same mean structure as {\sc CIM} and the variance structure is updated to the proposed structure of {\sc ImpCDM}, which we call {\sc CDM}.
Overall, we perform experiments using two datasets; reviews and profiles, with six different numbers of evaluators and four aggregation models; {\sc CIM}, {\sc ImpCIM}, {\sc CDM}, and {\sc ImpCDM}, and compute the correlation $20$ times using different initial values each time.
\color{black}

\subsection{Results and Discussion}
We first evaluate the results \textcolor{black}{based on} the review datasets.
Table~\ref{review_correlation_table} presents the mean correlations for five to ten workers computed over 20 trials.
\color{black}
In the potential quality, we observe that \textcolor{black}{the} {\sc ImpCDM} model with both proposed structures consistently outperforms {\sc CIM} in all the settings.
{\sc ImpCIM} also shows \textcolor{black}{a performance similar to that of} {\sc ImpCDM} while {\sc CDM} fails to improve \textcolor{black}{the} correlation from {\sc CIM} results.
This implies that our proposed mean structure which introduces workers' impression of evaluation targets mainly contributes to mitigate inter-criteria biases for potential quality estimation\textcolor{black}{.}
\textcolor{black}{Incorporating} it with our proposed variance structure which assumes constant variance among different criteria could lead to further improvement of the performance.

In the criteria quality, models with proposed structures also outperforms {\sc CIM}, but employing both structures is not the best choice.
{\sc ImpCDM} still shows the highest score among \textcolor{black}{the} four models mainly in ``coherence.'' 
\textcolor{black}{Alternatively}, {\sc CDM} works well in ``writing style'' criterion, and {\sc ImpCIM} performs the best in ``organization,'' `readability` and ``overall.''
{\sc ImpCIM} is not the best model in ``coherence,'' ``writing style'' and ``readability,'' but it consistently achieves better performance close to the best among \textcolor{black}{the} four aggregation models.
This reveals that our proposed mean structure helps to \textcolor{black}{model} inference to be robust against the inter-criteria biases for estimating criteria quality, but proposed variance structure has instability in performance and depends on the target criteria.

Next, we investigate the results \textcolor{black}{based on} the profile datasets. 
Table~\ref{profile_correlation_table} presents the mean correlations for five to ten workers computed over 20 trials.
\color{black}
Similar to the results on the review datasets, {\sc ImpCDM} achieves better prediction performance than {\sc CIM} in potential quality in most settings.
In contrast, {\sc CDM} and {\sc ImpCIM} show almost equal or lower performance than {\sc CIM}.
As in the case of the review datasets, incorporating our proposed mean structure with \textcolor{black}{the} proposed variance structure is effective in estimating the potential quality using simultaneous multi-criteria assessment.

In the criteria quality, proposed structures show more unstable performance than the results on the review datasets.
While {\sc CDM}, {\sc ImpCIM} or {\sc ImpCDM} outperform {\sc CIM} in most cases, {\sc ImpCDM} frequently fails to reach the highest performance among \textcolor{black}{the} four aggregation models.
We assume this is the limitation of our model extension, where we treat each criterion equally and independently.

In summary, our proposed aggregation models with the two proposed structural attachments alleviate the inter-criteria cognitive biases. 
In particular, the experimental results show the effectiveness of the proposed extensions in estimating the potential quality.
However, there is still room for improvement in the model design. 
While incorporating both proposed mean and variance structures improves prediction performance, their effects are limited under certain conditions, i.e., types of evaluation tasks \textcolor{black}{and} criteria of interest\textcolor{black}{.}
We could improve the prediction performance by incorporating some specific knowledge into the assessment generation process or introducing semantic relation between criteria.

\section{Conclusion}\label{conclusion_section}
In this study, we investigated the inter-criteria cognitive biases in multi-criteria \textcolor{black}{assessments} in crowdsourcing. 
We prepared two types of evaluation targets and collected responses for quality \textcolor{black}{assessments} in five criteria. 
Our exploratory analysis on our two response datasets revealed the existence of inter-criteria biases and how they affect worker ratings in terms of mean and variance.
Based on the observation, we proposed two model structures to incorporate the inter-criteria relations into the na\"ive aggregation model. 

\color{black}
We estimated true quality of \textcolor{black}{the} evaluation target using the na\"ive aggregation model and its extension with two \textcolor{black}{proposed} model structures and compared their prediction performance.
Our experimental results showed that the proposed model successfully made more accurate predictions than the baseline model, \textcolor{black}{specifically} in estimating the potential quality.
However, in estimating \textcolor{black}{the} criteria quality, \textcolor{black}{the} proposed variance structure failed to contribute to \textcolor{black}{the} performance improvement as much as \textcolor{black}{the} proposed mean structure.
The stability of inference of proposed models is particularly \textcolor{black}{reduced} when using profile datasets, which implies \textcolor{black}{that the proposed model are not robust to} various data conditions.
\color{black}

\color{black}
The contributions of this \textcolor{black}{study} are\textcolor{black}{:} (i) analysis based \textcolor{black}{on} identification of inter-criteria cognitive biases in multi-criteria assessment, (ii) construction of real multi-criteria assessment datasets using crowdsourcing for bias demonstration and model evaluation, and (iii) proposition of bias mitigation model structures and experimental investigation of their performance \textcolor{black}{in} several conditions.
One of the main limitation of our study is the prediction performance of criteria quality.
As demonstrated in \textcolor{black}{the} experimental results, our proposed aggregation model fails to consistently predict criteria quality.
We assume this is due to the design of the model that ignores relations among criteria.
In multi-criteria \textcolor{black}{assessments}, while criteria have different perspectives, they might share some common factors, e.g., differences in impressions of workers.
Therefore, grouping criteria into several clusters is one possible extension to represent such commonality for better prediction performance.
\color{black}

\color{black}
There are several future directions for research on mitigating inter-criteria cognitive biases in crowdsourcing.
From a practical \textcolor{black}{viewpoint}, it would be \textcolor{black}{an} interesting and important research direction to explore task designs of multi-criteria \textcolor{black}{assessments}.
\textcolor{black}{Another} direction is to expand the Likert scale more broadly, such as a hundred-point scale, which might \textcolor{black}{clarify} the tendency of ratings and the effects of inter-criteria biases\textcolor{black}{.}
Moreover, the evaluation \textcolor{black}{objective forms} \textcolor{black}{are} also a perspective \textcolor{black}{to} be explored.
For example, evaluating images in multiple criteria gives perceptions to workers different from that of texts, and could result in different biases.

\section*{Acknowledgments}
This work was supported by JST CREST Grant Number JPMJCR21D1.

\bibliography{reference}

\begin{thebibliography}{38}
\expandafter\ifx\csname natexlab\endcsname\relax\def\natexlab#1{#1}\fi
\providecommand{\url}[1]{\texttt{#1}}
\providecommand{\href}[2]{#2}
\providecommand{\path}[1]{#1}
\providecommand{\DOIprefix}{doi:}
\providecommand{\ArXivprefix}{arXiv:}
\providecommand{\URLprefix}{URL: }
\providecommand{\Pubmedprefix}{pmid:}
\providecommand{\doi}[1]{\href{http://dx.doi.org/#1}{\path{#1}}}
\providecommand{\Pubmed}[1]{\href{pmid:#1}{\path{#1}}}
\providecommand{\bibinfo}[2]{#2}
\ifx\xfnm\relax \def\xfnm[#1]{\unskip,\space#1}\fi
\bibitem[{Aipe \& Gadiraju(2018)}]{aipe2018}
\bibinfo{author}{Aipe, A.}, \& \bibinfo{author}{Gadiraju, U.}
  (\bibinfo{year}{2018}).
\newblock \bibinfo{title}{Similarhits: Revealing the role of task similarity in
  microtask crowdsourcing}.
\newblock In {\it \bibinfo{booktitle}{Proceedings of the 29th on Hypertext and
  Social Media}\/} (pp. \bibinfo{pages}{115--122}).
\bibitem[{Baba \& Kashima(2013)}]{baba2013}
\bibinfo{author}{Baba, Y.}, \& \bibinfo{author}{Kashima, H.}
  (\bibinfo{year}{2013}).
\newblock \bibinfo{title}{Statistical quality estimation for general
  crowdsourcing tasks}.
\newblock In {\it \bibinfo{booktitle}{Proceedings of the 19th ACM SIGKDD
  international conference on Knowledge discovery and data mining (KDD)}\/}
  (pp. \bibinfo{pages}{554--562}).
\bibitem[{Balzer \& Sulsky(1992)}]{balzer1992}
\bibinfo{author}{Balzer, W.~K.}, \& \bibinfo{author}{Sulsky, L.~M.}
  (\bibinfo{year}{1992}).
\newblock \bibinfo{title}{Halo and performance appraisal research: A critical
  examination}.
\newblock {\it \bibinfo{journal}{Applied Psychology}\/},  {\it
  \bibinfo{volume}{6}\/}, \bibinfo{pages}{975--985}.
\bibitem[{Barbera et~al.(2020)Barbera, Roitero, Demartini, Mizzaro \&
  Spina}]{barbera2020}
\bibinfo{author}{Barbera, D.~L.}, \bibinfo{author}{Roitero, K.},
  \bibinfo{author}{Demartini, G.}, \bibinfo{author}{Mizzaro, S.}, \&
  \bibinfo{author}{Spina, D.} (\bibinfo{year}{2020}).
\newblock \bibinfo{title}{Crowdsourcing truthfulness: The impact of judgment
  scale and assessor bias}.
\newblock In {\it \bibinfo{booktitle}{Proceedings of European Conference on
  Information Retrieval}\/} (pp. \bibinfo{pages}{207--214}).
\bibitem[{Biel \& Gatica-Perez(2014)}]{biel2014}
\bibinfo{author}{Biel, J.-I.}, \& \bibinfo{author}{Gatica-Perez, D.}
  (\bibinfo{year}{2014}).
\newblock \bibinfo{title}{Mining crowdsourced first impressions in online
  social video}.
\newblock {\it \bibinfo{journal}{IEEE Transactions on Multimedia}\/},  {\it
  \bibinfo{volume}{16}\/}, \bibinfo{pages}{2062--2074}.
\bibitem[{Brunner \& Munzel(2000)}]{brunner2000}
\bibinfo{author}{Brunner, E.}, \& \bibinfo{author}{Munzel, U.}
  (\bibinfo{year}{2000}).
\newblock \bibinfo{title}{The nonparametric behrens‐fisher problem:
  Asymptotic theory and a small‐sample approximation}.
\newblock {\it \bibinfo{journal}{Biometrical Journal}\/},  {\it
  \bibinfo{volume}{42}\/}, \bibinfo{pages}{17--25}.
\bibitem[{Cai et~al.(2016)Cai, Iqbal \& Teevan}]{cai2016}
\bibinfo{author}{Cai, C.~J.}, \bibinfo{author}{Iqbal, S.~T.}, \&
  \bibinfo{author}{Teevan, J.} (\bibinfo{year}{2016}).
\newblock \bibinfo{title}{Chain reactions: The impact of order on microtask
  chains}.
\newblock In {\it \bibinfo{booktitle}{Proceedings of the 2016 CHI Conference on
  Human Factors in Computing Systems (CHI)}\/} (pp.
  \bibinfo{pages}{3143--3154}).
\bibitem[{Coscia \& Rossi(2020)}]{coscia2020}
\bibinfo{author}{Coscia, M.}, \& \bibinfo{author}{Rossi, L.}
  (\bibinfo{year}{2020}).
\newblock \bibinfo{title}{Distortions of political bias in crowdsourced
  misinformation flagging}.
\newblock {\it \bibinfo{journal}{Journal of the Royal Society Interface}\/},
  {\it \bibinfo{volume}{17}\/}, \bibinfo{pages}{20200020}.
\bibitem[{Dawid \& Skene(1979)}]{dawid1979}
\bibinfo{author}{Dawid, A.~P.}, \& \bibinfo{author}{Skene, A.}
  (\bibinfo{year}{1979}).
\newblock \bibinfo{title}{Maximum likelihood estimation of observer error-rates
  using the em algorithm}.
\newblock {\it \bibinfo{journal}{Applied Statistics}\/},  {\it
  \bibinfo{volume}{28}\/}, \bibinfo{pages}{20--28}.
\bibitem[{DeCoths(1977)}]{decoths1977}
\bibinfo{author}{DeCoths, T.~A.} (\bibinfo{year}{1977}).
\newblock \bibinfo{title}{An analysis of the external validity and applied
  relevance of three rating formats}.
\newblock {\it \bibinfo{journal}{Organizational Behavior and Human
  Performance}\/},  {\it \bibinfo{volume}{19}\/}, \bibinfo{pages}{247--266}.
\bibitem[{Eickhoff(2018)}]{eickhoff2018}
\bibinfo{author}{Eickhoff, C.} (\bibinfo{year}{2018}).
\newblock \bibinfo{title}{Cognitive biases in crowdsourcing}.
\newblock In {\it \bibinfo{booktitle}{Proceedings of the Eleventh ACM
  International Conference on Web Search and Data Mining (WSDM)}\/} (pp.
  \bibinfo{pages}{162--170}).
\bibitem[{Eickhoff \& de~Vries(2013)}]{eickhoff2013}
\bibinfo{author}{Eickhoff, C.}, \& \bibinfo{author}{de~Vries, A.~P.}
  (\bibinfo{year}{2013}).
\newblock \bibinfo{title}{Increasing cheat robustness of crowdsourcing tasks}.
\newblock {\it \bibinfo{journal}{Information Retrieval}\/},  {\it
  \bibinfo{volume}{16}\/}, \bibinfo{pages}{121--137}.
\bibitem[{Gadiraju et~al.(2017{\natexlab{a}})Gadiraju, Fetahu, Kawase, Siehndel
  \& Dietze}]{gadiraju2017self}
\bibinfo{author}{Gadiraju, U.}, \bibinfo{author}{Fetahu, B.},
  \bibinfo{author}{Kawase, R.}, \bibinfo{author}{Siehndel, P.}, \&
  \bibinfo{author}{Dietze, S.} (\bibinfo{year}{2017}{\natexlab{a}}).
\newblock \bibinfo{title}{Using worker self-assessments for competence-based
  pre-selection in crowdsourcing microtasks}.
\newblock {\it \bibinfo{journal}{ACM Transactions on Computer-Human Interaction
  (TOCHI)}\/},  {\it \bibinfo{volume}{24}\/}, \bibinfo{pages}{30}.
\bibitem[{Gadiraju et~al.(2017{\natexlab{b}})Gadiraju, Yang \&
  Bozzon}]{gadiraju2017}
\bibinfo{author}{Gadiraju, U.}, \bibinfo{author}{Yang, J.}, \&
  \bibinfo{author}{Bozzon, A.} (\bibinfo{year}{2017}{\natexlab{b}}).
\newblock \bibinfo{title}{Clarity is a worthwhile quality: On the role of task
  clarity in microtask crowdsourcing}.
\newblock In {\it \bibinfo{booktitle}{Proceeding of the 28th ACM Conference on
  Hypertext and Social Media (HT)}\/}.
\bibitem[{Hoyt(2000)}]{hoyt2000}
\bibinfo{author}{Hoyt, W.~T.} (\bibinfo{year}{2000}).
\newblock \bibinfo{title}{Rater bias in psychological research: When is it a
  problem and what can we do about it?}
\newblock {\it \bibinfo{journal}{Psychological Methods}\/},  {\it
  \bibinfo{volume}{5}\/}, \bibinfo{pages}{64--86}.
\bibitem[{Hube et~al.(2019)Hube, Fetahu \& Gadiraju}]{hube2019}
\bibinfo{author}{Hube, C.}, \bibinfo{author}{Fetahu, B.}, \&
  \bibinfo{author}{Gadiraju, U.} (\bibinfo{year}{2019}).
\newblock \bibinfo{title}{Understanding and mitigating worker biases in the
  crowdsourced collection of subjective judgments}.
\newblock In {\it \bibinfo{booktitle}{Proceedings of the 2019 CHI Conference on
  Human Factors in Computing Systems (CHI)}\/} (pp. \bibinfo{pages}{1--12}).
\bibitem[{Kamar et~al.(2015)Kamar, Kappor \& Horvitz}]{kamar2015}
\bibinfo{author}{Kamar, E.}, \bibinfo{author}{Kappor, A.}, \&
  \bibinfo{author}{Horvitz, E.} (\bibinfo{year}{2015}).
\newblock \bibinfo{title}{Identifying and accounting for task-dependent bias in
  crowdsourcing}.
\newblock In {\it \bibinfo{booktitle}{Proceedings of the Third AAAI Conference
  on Human Computation and Crowdsourcing (HCOMP)}\/} (pp.
  \bibinfo{pages}{92--101}).
\bibitem[{Kulkarni et~al.(2014)Kulkarni, Socher, Bernstein \&
  Klemmer}]{kulkarni2014}
\bibinfo{author}{Kulkarni, C.~E.}, \bibinfo{author}{Socher, R.},
  \bibinfo{author}{Bernstein, M.~S.}, \& \bibinfo{author}{Klemmer, S.~R.}
  (\bibinfo{year}{2014}).
\newblock \bibinfo{title}{Scaling short-answer grading by combining peer
  assessment with algorithmic scoring}.
\newblock In {\it \bibinfo{booktitle}{Proceedings of the first ACM conference
  on Learning @ scale conference (L@S)}\/} (pp. \bibinfo{pages}{99--108}).
\bibitem[{Lakkaraju et~al.(2015)Lakkaraju, Leskovec, Kleinberg \&
  Mullainathan}]{lakkaraju2015}
\bibinfo{author}{Lakkaraju, H.}, \bibinfo{author}{Leskovec, J.},
  \bibinfo{author}{Kleinberg, J.}, \& \bibinfo{author}{Mullainathan, S.}
  (\bibinfo{year}{2015}).
\newblock \bibinfo{title}{A bayesian framework for modeling human evaluations}.
\newblock In {\it \bibinfo{booktitle}{Proceedings of the 2015 SIAM
  International Conference on Data Mining}\/} (pp. \bibinfo{pages}{181--189}).
\bibitem[{Lester \& Porter(1997)}]{Lester1997}
\bibinfo{author}{Lester, J.~C.}, \& \bibinfo{author}{Porter, B.~W.}
  (\bibinfo{year}{1997}).
\newblock \bibinfo{title}{Developing and empirically evaluating robust
  explanation generators: The {KNIGHT} experiments}.
\newblock {\it \bibinfo{journal}{Computational Linguistics}\/},  {\it
  \bibinfo{volume}{23}\/}, \bibinfo{pages}{65--101}.
\bibitem[{Li et~al.(2019)Li, Rubinstein \& Cohn}]{Li2019-EBCC}
\bibinfo{author}{Li, Y.}, \bibinfo{author}{Rubinstein, B.}, \&
  \bibinfo{author}{Cohn, T.} (\bibinfo{year}{2019}).
\newblock \bibinfo{title}{Exploiting worker correlation for label aggregation
  in crowdsourcing}.
\newblock In {\it \bibinfo{booktitle}{Proceedings of the International
  Conference on Machine Learning (ICML)}\/} (pp. \bibinfo{pages}{3886--3895}).
\newblock volume~\bibinfo{volume}{97}.
\bibitem[{Manouselis \& Costopoulou(2007)}]{manouselis2007}
\bibinfo{author}{Manouselis, N.}, \& \bibinfo{author}{Costopoulou, C.}
  (\bibinfo{year}{2007}).
\newblock \bibinfo{title}{Analysis and classification of multi-criteria
  recommender systems}.
\newblock {\it \bibinfo{journal}{World Wide Web}\/},  {\it
  \bibinfo{volume}{10}\/}, \bibinfo{pages}{415--441}.
\bibitem[{Mellish \& Dale(1998)}]{Mellish1998}
\bibinfo{author}{Mellish, C.}, \& \bibinfo{author}{Dale, R.}
  (\bibinfo{year}{1998}).
\newblock \bibinfo{title}{Evaluation in the context of natural language
  generation}.
\newblock {\it \bibinfo{journal}{Computer Speech \& Language}\/},  {\it
  \bibinfo{volume}{12}\/}, \bibinfo{pages}{349--373}.
\bibitem[{Newell \& Ruths(2016)}]{newell2016}
\bibinfo{author}{Newell, E.}, \& \bibinfo{author}{Ruths, D.}
  (\bibinfo{year}{2016}).
\newblock \bibinfo{title}{How one microtask affects another}.
\newblock In {\it \bibinfo{booktitle}{Proceedings of the 2016 CHI Conference on
  Human Factors in Computing Systems (CHI)}\/} (pp.
  \bibinfo{pages}{3155--3166}).
\bibitem[{Nguyen et~al.(2016)Nguyen, Halpern, Wallace \& Lease}]{nguyen2016}
\bibinfo{author}{Nguyen, A.~T.}, \bibinfo{author}{Halpern, M.},
  \bibinfo{author}{Wallace, B.~C.}, \& \bibinfo{author}{Lease, M.}
  (\bibinfo{year}{2016}).
\newblock \bibinfo{title}{Probabilistic modeling for crowdsourcing
  partially-subjective ratings}.
\newblock In {\it \bibinfo{booktitle}{Proceedings of the Fourth AAAI Conference
  on Human Computation and Crowdsourcing (HCOMP)}\/} (pp.
  \bibinfo{pages}{149--158}).
\bibitem[{Raykar \& Yu(2011)}]{raykar2011}
\bibinfo{author}{Raykar, V.~C.}, \& \bibinfo{author}{Yu, S.}
  (\bibinfo{year}{2011}).
\newblock \bibinfo{title}{Ranking annotators for crowdsourced labeling tasks}.
\newblock In {\it \bibinfo{booktitle}{Proceedings of the 24th International
  Conference on Neural Information Processing Systems (NIPS)}\/} (pp.
  \bibinfo{pages}{1809--1817}).
\bibitem[{Raykar et~al.(2010)Raykar, Zhao, Valadez, Florin, Bogoni \&
  Moy}]{raykar2010}
\bibinfo{author}{Raykar, V.~C.}, \bibinfo{author}{Zhao, L.~H.},
  \bibinfo{author}{Valadez, G.~H.}, \bibinfo{author}{Florin, C.},
  \bibinfo{author}{Bogoni, L.}, \& \bibinfo{author}{Moy, L.}
  (\bibinfo{year}{2010}).
\newblock \bibinfo{title}{Learning from crowds}.
\newblock {\it \bibinfo{journal}{Journal of Machine Learning Research}\/},
  {\it \bibinfo{volume}{11}\/}, \bibinfo{pages}{1297--1322}.
\bibitem[{Saal et~al.(1980)Saal, Downey \& Lahey}]{saal1980}
\bibinfo{author}{Saal, F.~E.}, \bibinfo{author}{Downey, R.~G.}, \&
  \bibinfo{author}{Lahey, M.~A.} (\bibinfo{year}{1980}).
\newblock \bibinfo{title}{Rating the ratings: Assessing the psychometric
  quality of rating data}.
\newblock {\it \bibinfo{journal}{Psychological Bulletin}\/},  {\it
  \bibinfo{volume}{88}\/}, \bibinfo{pages}{413--428}.
\bibitem[{Salakhutdinov \& Mnih(2008)}]{salakhutdinov2008}
\bibinfo{author}{Salakhutdinov, R.}, \& \bibinfo{author}{Mnih, A.}
  (\bibinfo{year}{2008}).
\newblock \bibinfo{title}{Bayesian probabilistic matrix factorization using
  markov chain monte carlo}.
\newblock In {\it \bibinfo{booktitle}{Proceedings of the 25th International
  Conference on Machine Learning (ICML)}\/} (pp. \bibinfo{pages}{880--887}).
\bibitem[{Samejima(1969)}]{samejima1969}
\bibinfo{author}{Samejima, F.} (\bibinfo{year}{1969}).
\newblock \bibinfo{title}{Estimation of latent ability using a response pattern
  of graded scores}.
\newblock {\it \bibinfo{journal}{Psychometrika Supplement 1}\/},  {\it
  \bibinfo{volume}{34}\/}, \bibinfo{pages}{1--97}.
\bibitem[{Snow et~al.(2008)Snow, O'Connor, Jurafsky \& Ng}]{snow2008}
\bibinfo{author}{Snow, R.}, \bibinfo{author}{O'Connor, B.},
  \bibinfo{author}{Jurafsky, D.}, \& \bibinfo{author}{Ng, A.~Y.}
  (\bibinfo{year}{2008}).
\newblock \bibinfo{title}{Cheap and fast---but is it good? evaluating
  non-expert annotations for natural language tasks}.
\newblock In {\it \bibinfo{booktitle}{Proceedings of the 2008 Conference on
  Empitical Methods in Natural Language Processing (EMNLP)}\/} (pp.
  \bibinfo{pages}{254--263}).
\bibitem[{Thorndike(1920)}]{thorndike1920}
\bibinfo{author}{Thorndike, E.~L.} (\bibinfo{year}{1920}).
\newblock \bibinfo{title}{A constant error in psychological ratings}.
\newblock {\it \bibinfo{journal}{Applied Psychology}\/},  {\it
  \bibinfo{volume}{4}\/}, \bibinfo{pages}{25--29}.
\bibitem[{Venanzi et~al.(2014)Venanzi, Guiver, Kazai, Kohli \&
  Shokouhi}]{Venanzi2014}
\bibinfo{author}{Venanzi, M.}, \bibinfo{author}{Guiver, J.},
  \bibinfo{author}{Kazai, G.}, \bibinfo{author}{Kohli, P.}, \&
  \bibinfo{author}{Shokouhi, M.} (\bibinfo{year}{2014}).
\newblock \bibinfo{title}{Community-based bayesian aggregation models for
  crowdsourcing}.
\newblock In {\it \bibinfo{booktitle}{Proceedings of the International
  Conference on World wide web (WWW)}\/} (pp. \bibinfo{pages}{155--164}).
\bibitem[{Vougiouklis et~al.(2018)Vougiouklis, Elsahar, Kaffee, Gravier,
  Laforest, Hare \& Simperl}]{Vougiouklis2018NeuralWikipedian}
\bibinfo{author}{Vougiouklis, P.}, \bibinfo{author}{Elsahar, H.},
  \bibinfo{author}{Kaffee, L.-A.}, \bibinfo{author}{Gravier, C.},
  \bibinfo{author}{Laforest, F.}, \bibinfo{author}{Hare, J.}, \&
  \bibinfo{author}{Simperl, E.} (\bibinfo{year}{2018}).
\newblock \bibinfo{title}{Neural wikipedian: Generating textual summaries from
  knowledge base triples}.
\newblock {\it \bibinfo{journal}{Journal of Web Semantics}\/},  {\it
  \bibinfo{volume}{52}\/}, \bibinfo{pages}{1--15}.
\bibitem[{Welinder et~al.(2011)Welinder, Branson, Belongie \&
  Perona}]{welinder2010}
\bibinfo{author}{Welinder, P.}, \bibinfo{author}{Branson, S.},
  \bibinfo{author}{Belongie, S.}, \& \bibinfo{author}{Perona, P.}
  (\bibinfo{year}{2011}).
\newblock \bibinfo{title}{The multidimensional wisdom of crowds}.
\newblock In {\it \bibinfo{booktitle}{Proceedings of the 23rd International
  Conference on Neural Information Processing Systems (NIPS)}\/} (pp.
  \bibinfo{pages}{1809--1817}).
\newblock volume~\bibinfo{volume}{2}.
\bibitem[{Whitehill et~al.(2009)Whitehill, fan Wu, Bergsma, Movellan \&
  Ruvolo}]{whitehill2009}
\bibinfo{author}{Whitehill, J.}, \bibinfo{author}{fan Wu, T.},
  \bibinfo{author}{Bergsma, J.}, \bibinfo{author}{Movellan, J.~R.}, \&
  \bibinfo{author}{Ruvolo, P.~L.} (\bibinfo{year}{2009}).
\newblock \bibinfo{title}{Whose vote should count more: Optimal integration of
  labels from labelers of unknown expertise}.
\newblock In {\it \bibinfo{booktitle}{Proceedings of the 22nd International
  Conference on Neural Information Processing Systems (NIPS)}\/} (pp.
  \bibinfo{pages}{2035--2043}).
\bibitem[{Zhuang et~al.(2015)Zhuang, Parameswaran, Roth \& Han}]{zhuang2015kdd}
\bibinfo{author}{Zhuang, H.}, \bibinfo{author}{Parameswaran, A.},
  \bibinfo{author}{Roth, D.}, \& \bibinfo{author}{Han, J.}
  (\bibinfo{year}{2015}).
\newblock \bibinfo{title}{Debiasing crowdsourced batches}.
\newblock In {\it \bibinfo{booktitle}{Proceedings of the 21st ACM SIGKDD
  International Conference on Knowledge Discovery and Data Mining (KDD)}\/}
  (pp. \bibinfo{pages}{1593--1602}).
\bibitem[{Zhuang \& Young(2015)}]{zhuang2015wsdm}
\bibinfo{author}{Zhuang, H.}, \& \bibinfo{author}{Young, J.}
  (\bibinfo{year}{2015}).
\newblock \bibinfo{title}{Leveraging in-batch annotation bias for crowdsourced
  active learning}.
\newblock In {\it \bibinfo{booktitle}{Proceedings of the Eighth ACM
  International Conference on Web Search and Data Mining (WSDM)}\/} (pp.
  \bibinfo{pages}{243--252}).

\end{thebibliography}

\end{document}